# COMPLETE RESULTS FOR FIVE YEARS OF GNO SOLAR NEUTRINO OBSERVATIONS

GNO COLLABORATION

M. Altmann[a], M. Balata[b], P. Belli[c], E. Bellotti[d,x], R. Bernabei[c], E. Burkert[e], C. Cattadori[d], R. Cerulli[c], M. Chiarini[f], M. Cribier[g], S. d'Angelo[c], G. Del Re[f], K.H. Ebert[h], F. v. Feilitzsch[i], N. Ferrari[b], W. Hampel[e], F.X. Hartmann[e], E. Henrich[h], G. Heusser[e], F. Kaether[e], J. Kiko[e], T. Kirsten[e,y], T. Lachenmaier[i], J. Lanfranchi[i], M. Laubenstein[b], K. Lützenkirchen[j], K. Mayer[j], P. Moegel[e], D. Motta[e], S. Nisi[b], J. Oehm[e], L. Pandola[b,k], F. Petricca[a,d], W. Potzel[i], H. Richter[e], S. Schoenert[e], M. Wallenius[j], M. Wojcik[l], L. Zanotti[d]

[a] Max-Planck-Institut für Physik (Werner-Heisenberg-Institut), Föhringer Ring 6, D-80805 München, Germany
[b] INFN, Laboratori Nazionali del Gran Sasso (LNGS), S.S. 17/bis Km 18+910, I-67010 l'Aquila, Italy [1]
[c] Dipartimento di Fisica, Università di Roma 'Tor Vergata' e INFN, Sezione di Roma II, Via della Ricerca Scientifica, I-00133 Roma, Italy [1]
[d] Dipartimento di Fisica, Università di Milano 'La Bicocca' e INFN, Sezione di Milano, Via Emanueli, I-20126 Milano, Italy [1]
[e] Max-Planck-Institut für Kernphysik (MPIK), P.O.B. 103980, D-69029 Heidelberg, Germany[2,3]
[f] Dipartimento di Ingegneria Chimica e Materiali, Università dell'Aquila, Località Monteluco di Roio, l'Aquila, Italy [1]
[g] DAPNIA/Service de Physique des Particules, CEA Saclay, F-91191 Gif-sur-Yvette Cedex, France
[h] Institut für Technische Chemie, Forschungszentrum Karlsruhe (FZK), P.O.B. 3640, D-76021 Karlsruhe, Germany
[i] Physik-Department E15, Technische Universität München (TUM), James-Franck-Straße, D-85748 Garching, Germany [4]
[j] European Commission, Joint Research Centre, Institute for Transuranium Elements, P.O.B. 2340, D-76125 Karlsruhe, Germany
[k] Dipartimento di Fisica, Università dell'Aquila, Via Vetoio 1, I-67010 Coppito, L'Aquila, Italy
[l] Instiytut Fizyki, Uniwersytet Jagiellonski, ul. Reymonta 4, PL-30059 Krakow, Poland
[x] Spokesperson
[y] Corresponding author

[1] This work has been supported by Istituto Nazionale di Fisica Nucleare (INFN), Italy.
[2] This work has been supported by the German Bundesministerium für Bildung und Forschung (BMBF).
[3] This work has been generously supported by the Alfried Krupp von Bohlen und Halbach- Foundation, Germany.
[4] This work has been supported in part by the Deutsche Forschungsgemeinschaft DFG (Sonderforschungsbereich 375), the German Bundesministerium für Bildung und Forschung (BMBF), the Maier-Leibnitz-Laboratorium (Munich) and the EEC cryonetwork program under contract numbers ERBFM-RXCT980167 and HPRN-CT-2002-00322.

Corresponding author: Till A. Kirsten, Max-Planck-Institut für Kernphysik, P.O. Box 103980, D-69029 Heidelberg.
Phone: +49 (0) 6220 913098; Fax: +49 (0) 6220 913289; E-mail: till.kirsten@mpi-hd.mpg.de

to be published in Physics Letters B
submitted: January 16, 2005
accepted after revision: April 13, 2005

**Abstract**

We report the complete GNO solar neutrino results for the measuring periods GNO III, GNO II, and GNO I. The result for GNO III (last 15 solar runs) is [54.3 ± $^{9.9}_{9.3}$ (stat.) ± 2.3 (syst.)] SNU (1σ) or [54.3 ± $^{10.2}_{9.6}$ (incl. syst.)] SNU (1σ) with errors combined. The GNO experiment is now terminated after altogether 58 solar exposure runs that were performed between May 20, 1998 and April 9, 2003. The combined result for GNO (I+II+III) is [62.9 ± $^{5.5}_{5.3}$ (stat.) ± 2.5 (syst.)] SNU (1σ) or [62.9 ± $^{6.0}_{5.9}$] SNU (1σ) with errors combined in quadrature. Overall, gallium based solar observations at LNGS (first in GALLEX, later in GNO) lasted from May 14, 1991 through April 9, 2003. The joint result from 123 runs in GNO and GALLEX is [69.3 ± 5.5 (incl. syst.)] SNU (1σ). The distribution of the individual run results is consistent with the hypothesis of a neutrino flux that is constant in time. Implications from the data in particle- and astrophysics are reiterated.

## 1. Introduction

The Gallium Neutrino Observatory (GNO) experiment at the Laboratori Nazionali del Gran Sasso (LNGS) has recorded solar neutrinos with energies above 233 keV via the inverse EC reaction $^{71}$Ga ($\nu_e$, e$^-$)$^{71}$Ge in a 100-ton gallium chloride target (containing 30.3 tons of gallium) between May 1998 and April 2003 [1-3]. The data taking has now been terminated for external non-scientific reasons. Together with the preceding GALLEX observations, low energy solar neutrino recordings were acquired for more than a full solar cycle (1991-2003, with a break in 1997).

The GALLEX recordings (till 1997) established both the presence of pp-neutrinos [4] and a significant deficit (≈ 40 %) in the sub-MeV neutrino induced rate [5-7]. At that time, this was the strongest indication for neutrino transformations on the way between the solar core and the Earth, implying non-zero neutrino mass and non-standard physics [8-11]. The subsequent GNO observations have improved the quality of the data, added important restrictions on the presence of possible time variations, and substantially reduced the total error on the charged current reaction rate for pp-neutrinos as measured by the inverse EC reaction on gallium. Without radiochemical gallium detectors, the majority (93 %) of all solar neutrinos would still remain unobserved.

Gallium experiments at LNGS have now provided a long time record of low energy solar neutrinos and determined the bulk production rate with an accuracy of ± 5.5 SNU. This is based on 123 solar runs ("SR"), 65 from GALLEX and 58 from GNO. Results for the first 19 GNO runs (GNO I) were published in [1], here we report the full data for SR20 – SR43 (GNO II) and for SR44 – SR58 (GNO III).

The result and its precision will remain without competition from eventually upcoming low-threshold real-time experiments for many years to come. Recorded is a fundamental astrophysical quantity, the neutrino luminosity of the Sun. Lack of these basic data would have a very negative impact on the interpretation of the results from forthcoming second generation (real-time and low threshold) solar neutrino experiments such as BOREXINO (aiming for $^7$Be neutrino detection). In the astrophysical context, the gallium results shed light also on the individual contributions of the PPI, PPII and CNO cycles to the solar luminosity and on the agreement of the energy production derived from the photon- and from the neutrino luminosity respectively.

The plan of this article is as follows: In sect. 2 we describe some new experimental aspects and summarize the run characteristics. Data evaluation as well as the complete results and their errors are presented in sect. 3. In sect. 4 we discuss the internal consistency of all GNO and GALLEX data and the analysis for possible time variations of the neutrino signal. Sect. 5 is devoted to the interpretation of the results in the context of particle physics (5.1) and of astrophysics (5.2).

## 2. Experimental

The basic procedures in performing GNO extraction runs have been described before [1,3-7,12]. Table 1 summarizes the GNO run characteristics. The operating periods defined as GNO II and GNO III comprised 24 and 15 solar runs respectively. Consecutive to the earlier 19 GNO I runs (results given in [1]), the new runs are labeled SR20–SR43 (GNO II) and SR44-SR58 (GNO III). The standard exposure time for solar runs was 4 weeks, with minor deviations due to practical reasons (Table 1, 5$^{th}$ column).

### 2.1 Ge extraction yields

Ge extraction yields were monitored with added Ge carriers ($\approx$ 1 mg per run). The stable isotopes $^{70}$Ge, $^{72}$Ge, $^{74}$Ge, and $^{76}$Ge were used alternately (Table 1, 6$^{th}$ column). The Ge recovery yields are listed in column 7 of Table 1. The yield refers to the percentage of initially added germanium that actually ended up in the counter. The complement of 100 % is the sum of the Ge that escaped extraction from the target and of the unavoidable losses that occur during the conversion of GeCl$_4$ into the counting gas component GeH$_4$ (germane) and in the counter filling procedure. Yields range from 91.2 % to 98.7 %, with a mean value of 95.7 %.

Compared to the standard Ge desorption conditions of GALLEX, desorption time and desorption gas volumes for GNO have been reduced from 12 to 9 hours and from 2500 to 1700 m$^3$ (at 20°C and 0.9 bar). This simplified the operating schedule at the expense of a slightly less efficient Ge desorption. The un-desorbed Ge-isotope carrier fraction remaining in the tank is expected to increase from about 0.2 % to values between 1 and 2 %. The fact that a homogeneously distributed Ge hold-back carrier level

**Table 1:** GNO - run characteristics

| Type[a] | Run# | Exposure time Start | Exposure time End | Duration [d] | Carrier[b] | Ge yield[c] [%] | Label[d] | Counting efficiency[e] used[%] | Counting efficiency[e] old [%] | Counting time [d] |
|---|---|---|---|---|---|---|---|---|---|---|
| **GNO I** | | | | | | | | | | |
| SR01 | EX03 | 20/05/98 | 17/06/98 | 28 | 72 | 96.1 | SC 138 | 80.2 ± 1.2 | 81.4 ± 3.5 | 179.6 |
| SR02 | EX04 | 17/06/98 | 22/07/98 | 35 | 74 | 93.5 | Fe118* | 74.3 ± 3.5 | 74.3 ± 3.5 | 173.5 |
| SR03 | EX05 | 22/07/98 | 26/08/98 | 35 | 76 | 95.1 | Si114* | 75.5 ± 3.5 | 75.5 ± 3.5 | 162.3 |
| SR04 | EX06 | 26/08/98 | 23/09/98 | 28 | 70 | 97.9 | Si 113* | 78.1 ± 3.5 | 78.1 ± 3.5 | 138.3 |
| SR05 | EX07 | 23/09/98 | 21/10/98 | 28 | 72 | 94.6 | FC093 | 79.9 ± 0.7 | 80.0 ± 3.5 | 137.7 |
| SR06 | EX08 | 21/10/98 | 18/11/98 | 28 | 74 | 94.5 | Si108 | 75.3 ± 0.7 | 77.9 ± 3.5 | 166.5 |
| SR07 | EX09 | 18/11/98 | 16/12/98 | 28 | 76 | 94.4 | SC136 | 80.9 ± 0.8 | 82.3 ± 3.5 | 180.6 |
| SR08 | EX10 | 16/12/98 | 13/01/99 | 28 | 70 | 96.8 | FC102 | 78.9 ± 0.9 | 81.4 ± 3.5 | 179.4 |
| SR09 | EX11 | 13/01/99 | 10/02/99 | 28 | 72 | 95.8 | SC135 | 82.3 ± 1.1 | 82.3 ± 1.1 | 194.7 |
| SR10 | EX13 | 10/03/99 | 14/04/99 | 35 | 76 | 94.6 | SC139 | 80.3 ± 1.0 | 81.3 ± 3.5 | 187.4 |
| SR11 | EX14 | 14/04/99 | 19/05/99 | 35 | 70 | 96.4 | Fe039 | 74.8 ± 0.9 | 76.4 ± 2.1 | 180.4 |
| SR12 | EX15 | 19/05/99 | 16/06/99 | 28 | 72 | 95.7 | Fe043 | 76.7 ± 0.8 | 76.7 ± 2.1 | 179.4 |
| SR13 | EX16 | 16/06/99 | 28/07/99 | 42 | 74 | 94.4 | SC136 | 80.9 ± 0.8 | 82.3 ± 3.5 | 165.8 |
| SR14 | EX17 | 28/07/99 | 25/08/99 | 28 | 76 | 96.0 | FC093 | 79.9 ± 0.7 | 80.0 ± 3.5 | 167.6 |
| SR15 | EX18 | 25/08/99 | 22/09/99 | 28 | 70 | 97.1 | FC102 | 78.9 ± 0.9 | 81.4 ± 3.5 | 165.6 |
| SR16 | EX19 | 22/09/99 | 20/10/99 | 28 | 72 | 95.7 | Si113* | 78.1 ± 3.5 | 78.1 ± 3.5 | 165.4 |
| SR17 | EX20 | 20/10/99 | 17/11/99 | 28 | 74 | 95.6 | SC139 | 80.3 ± 1.0 | 81.3 ± 3.5 | 166.5 |
| SR18 | EX21 | 17/11/99 | 14/12/99 | 27 | 76 | 94.7 | Fe039 | 74.8 ± 0.9 | 76.4 ± 2.1 | 166.6 |
| SR19 | EX22 | 14/12/99 | 12/01/00 | 29 | 70 | 91.5 | Si106 | 74.8 ± 1.1 | 77.1 ± 3.5 | 166.5 |
| **GNO II** | | | | | | | | | | |
| SR20 | EX24 | 13/01/00 | 09/02/00 | 27 | 72 | 91.2 | Si108 | 75.3 ± 0.7 | 77.9 ± 3.5 | 164.6 |
| SR21 | EX25 | 09/02/00 | 08/03/00 | 28 | 74 | 94.8 | FC093 | 79.9 ± 0.7 | 80.0 ± 3.5 | 167.8 |
| SR22 | EX26 | 08/03/00 | 05/04/00 | 28 | 76 | 97.7 | FC174 | 78.8 ± 0.9 | 81.7 ± 3.5 | 166.6 |
| SR23 | EX28 | 06/04/00 | 03/05/00 | 27 | 70 | 94.7 | SC136 | 80.9 ± 0.8 | 82.3 ± 3.5 | 167.8 |
| SR24 | EX29 | 03/05/00 | 31/05/00 | 28 | 72 | 93.0 | Fe039 | 74.8 ± 0.9 | 76.4 ± 2.1 | 165.8 |
| SR25 | EX32 | 29/06/00 | 26/07/00 | 27 | 74 | 93.9 | Si106 | 74.8 ± 1.1 | 77.1 ± 3.5 | 167.7 |
| SR26 | EX33 | 26/07/00 | 23/08/00 | 28 | 76 | 94.1 | SC138 | 80.2 ± 1.2 | 81.4 ± 3.5 | 167.8 |
| SR27 | EX34 | 23/08/00 | 20/09/00 | 28 | 70 | 97.2 | Si108 | 75.3 ± 0.7 | 77.9 ± 3.5 | 166.6 |
| SR28 | EX36 | 21/09/00 | 18/10/00 | 27 | 72 | 95.4 | FC174 | 78.8 ± 0.9 | 81.7 ± 3.5 | 166.6 |
| SR29 | EX37 | 18/10/00 | 15/11/00 | 28 | 74 | 93.9 | FC102 | 78.9 ± 0.9 | 81.4 ± 3.5 | 167.5 |
| SR30 | EX38 | 15/11/00 | 12/12/00 | 27 | 76 | 98.1 | SC136 | 80.9 ± 0.8 | 82.3 ± 3.5 | 166.5 |
| SR31 | EX40 | 13/12/00 | 10/01/01 | 27 | 70 | 96.2 | FC126 | 78.5 ± 0.6 | 75.4 ± 3.5 | 163.3 |
| SR32 | EX41 | 10/10/01 | 07/02/01 | 28 | 72 | 95.0 | SC139 | 80.3 ± 1.0 | 81.3 ± 3.5 | 165.6 |
| SR33 | EX42 | 07/02/01 | 07/03/01 | 28 | 74 | 97.8 | Si106 | 74.8 ± 1.1 | 77.1 ± 3.5 | 166.7 |
| SR34 | EX44 | 08/03/01 | 04/04/01 | 27 | 76 | 92.3 | FC093 | 79.9 ± 0.7 | 80.0 ± 3.5 | 167.8 |
| SR35 | EX45 | 04/04/01 | 03/05/01 | 29 | 70 | 95.1 | SC138 | 80.2 ± 1.2 | 81.4 ± 3.5 | 166.8 |
| SR36 | EX46 | 03/05/01 | 30/05/01 | 27 | 72 | 96.5 | Fe039 | 74.8 ± 0.9 | 76.4 ± 2.1 | 160.9 |
| SR37 | EX48 | 31/05/01 | 27/06/01 | 27 | 74 | 95.3 | FC126 | 78.5 ± 0.6 | 75.4 ± 3.5 | 166.7 |
| SR38 | EX49 | 27/06/01 | 25/07/01 | 28 | 76 | 95.3 | SC151 | 73.9 ± 1.1 | 75.2 ± 3.5 | 166.8 |
| SR39 | EX50 | 25/07/01 | 22/08/01 | 28 | 70 | 98.7 | SC150 | 78.9 ± 1.2 | 75.0 ± 3.5 | 167.2 |
| SR40 | EX 51 | 22/08/01 | 19/09/01 | 28 | 72 | 96.2 | Fe112* | 72.9 ± 3.5 | 72.9 ± 3.5 | 167.9 |
| SR41 | EX 53 | 20/09/01 | 17/10/01 | 27 | 74 | 94.1 | SC139 | 80.3 ± 1.0 | 81.3 ± 3.5 | 174.6 |
| SR42 | EX 54 | 17/10/01 | 14/11/01 | 28 | 76 | 97.2 | Si108 | 75.3 ± 0.7 | 77.9 ± 3.5 | 174.7 |
| SR43 | EX 57 | 13/12/01 | 08/01/02 | 26 | 70 | 96.2 | SC156* | 80.6 ± 3.5 | 80.6 ± 3.5 | 172.5 |
| **GNO III** | | | | | | | | | | |
| SR44 | EX58 | 09/01/02 | 06/02/02 | 28 | 72 | 98.7 | FC126 | 78.5 ± 0.6 | | 174.7 |
| SR45 | EX59 | 06/02/02 | 06/03/02 | 28 | 74 | 96.1 | SC151 | 73.9 ± 1.1 | | 172.5 |
| SR46 | EX61 | 07/03/02 | 10/04/02 | 34 | 76 | 94.8 | SC136 | 80.9 ± 0.8 | | 183.0 |
| SR47 | EX62 | 10/04/02 | 08/05/02 | 28 | 70 | 97.3 | Si106 | 74.8 ± 1.1 | | 165.6 |
| SR48 | EX63 | 08/05/02 | 05/06/02 | 28 | 72 | 95.1 | Fe112* | 72.9 ± 3.5 | | 166.7 |
| SR49 | EX65 | 06/06/02 | 03/07/02 | 27 | 74 | 93.8 | FC093 | 79.9 ± 0.7 | | 166.6 |
| SR50 | EX66 | 03/07/02 | 31/07/02 | 28 | 76 | 97.8 | FC174 | 78.8 ± 0.9 | | 167.9 |
| SR51 | EX67 | 31/07/02 | 28/08/02 | 28 | 70 | 93.6 | FC102 | 78.9 ± 0.9 | | 166.6 |
| SR52 | EX68 | 28/08/02 | 23/10/02 | 56 | 72 | 97.3 | SC136 | 80.9 ± 0.8 | | 165.9 |
| SR53 | EX69 | 23/10/02 | 20/11/02 | 28 | 74 | 96.4 | Si106 | 74.8 ± 1.1 | | 181.8 |
| SR54 | EX71 | 21/11/02 | 18/12/02 | 27 | 76 | 98.3 | SC150 | 78.9 ± 1.2 | | 243.0 |
| SR55 | EX72 | 18/12/02 | 15/01/03 | 28 | 70 | 97.3 | FC093 | 79.9 ± 0.7 | | 319.8 |
| SR56 | EX73 | 15/01/03 | 12/02/03 | 28 | 72 | 97.0 | FC174 | 78.8 ± 0.9 | | 312.5 |
| SR57 | EX75 | 13/02/03 | 12/03/03 | 27 | 74 | 98.3 | Si108 | 75.3 ± 0.7 | | 256.6 |
| SR58 | EX76 | 12/03/03 | 09/04/03 | 28 | 76 | 95.9 | SC136 | 80.9 ± 0.8 | | 227.8 |

[a] SR = solar neutrino run.
[b] 70, 72, 74, 76 indicate the use of carrier solutions enriched in $^{70}$Ge, $^{72}$Ge, $^{74}$Ge, $^{76}$Ge, respectively.
[c] Integral tank-to-counter yield of Ge-carriers. The combined error assigned to the uncertainties of the yields and of the target mass is 2.2 %.
[d] Counters have either iron (Fe) or silicon (Si) cathode. SC = silicon counter with shaped cathode, FC = iron counter with shaped cathode.
[e] Efficiency > 0.5 keV. For distinction between "used" and "old", see text.
* counter that is not absolutely calibrated (see sect. 2.2).

never drops below $\approx 10^{-13}$ mol per liter helps to exclude hypothetical $^{71}$Ge loss scenarios that would involve the carrying-in of non measurable ultra-low trace impurities below that level.

The unavoidable Ge carrier residue that remains un-desorbed in a run is desorbed during the next run. Its quantity can be determined by mass spectrometric analysis of the Ge recovered from the GeH$_4$ after counting. This is possible since we use in alternating sequence four different enriched stable Ge isotopes ($^{70}$Ge, $^{72}$Ge, $^{74}$Ge and $^{76}$Ge) as mentioned above. The measurements are done with a multi-collector inductively coupled plasma mass spectrometer (MC-ICP-MS; NU Instruments, Wrexham, UK). Until now, about one third of the GNO runs have been analyzed in this way.

Based on textbook principles of isotope dilution analysis, the measured mass spectra allow splitting the composition of the analyzed Ge into the three potential contributions: principal carrier, carry-over from the foregoing run, and Ge with natural isotopic composition (chemical contamination at trace level). The results shown in Table 2 are required for the precise determination of the chemical

**Table 2:** Mass-spectrometric analysis of Ge-samples (recovered from the counting gas) for decomposition into the three contributing components.

| Run # | Carrier [%] | Carry-over from previous run [%] | Third component (contaminant)[a] [%] |
|---|---|---|---|
| GNO I - EX01 * | 99.1 | 0.0 | 0.9 |
| GNO I - EX02 * | 99.6 | 0.4 | 0.0 |
| GNO I - SR01 * | 98.3 | 1.4 | 0.3 |
| GNO I - SR02 | 99.1 | 0.7 | 0.2 |
| GNO I - SR03 | 98.2 | 1.8 | 0.0 |
| GNO I - SR04 | 98.4 | 1.4 | 0.2 |
| GNO I - SR05 * | 91.1 | 1.5 | 7.4 ($^{74}$Ge) |
| GNO I - SR06 | 98.2 | 1.5 | 0.3 |
| GNO I - SR07 | 98.3 | 1.5 | 0.2 |
| GNO I - SR08 | 98.3 | 1.5 | 0.2 |
| GNO I - SR09 | 98.3 | 1.5 | 0.2 |
| GNO I - SR10 | 98.1 | 1.65 | 0.3 |
| GNO I - SR11 | 98.4 | 1.3 | 0.3 |
| GNO I - SR12 | 98.6 | 1.25 | 0.15 |
| GNO I - SR13 | 98.4 | 1.4 | 0.25 |
| GNO I - SR14 | 98.5 | 1.3 | 0.2 |
| GNO I - SR15 * | 97.0 | 1.55 | 1.45 ($^{72}$Ge) |
| GNO I - SR16 * | 90.5 | 1.5 | 8.0 ($^{76}$Ge) |
| GNO I - SR17 * | 94.1 | 5.9 | 0.0 |
| GNO I - SR18 * | 93.5 | 5.8 | 0.7 |
| GNO I - SR19 * | 94.6 | 2.0 | 3.4 ($^{74}$Ge) |
| GNO II - BL01 * | 92.8 | 2.5 | 4.7 ($^{76}$Ge) |
| GNO II - SR20 | 98.8 | 1.1 | 0.1 |
| GNO II - SR21 | 97.0 | 1.5 | 1.5 |
| GNO II - SR22 | 97.2 | 1.7 | 1.15 |
| GNO II - BL02 | 98.0 | 1.25 | 0.75 |
| Mean[b] | 98.23 ± 0.53 | 1.4 ± 0.26 | 0.38 ± 0.41 |

[a] Ge with natural isotopic composition, if not indicated otherwise.
[b] does not include samples marked * due to known irregularities in the counting gas preservation and recovery or in the Ge preparation procedure for mass spectrometric analysis.

$^{71}$Ge recovery yields in the extraction runs. A first estimate from the samples analyzed so far is (1.4 ± 0.3) % for the residual Ge from the foregoing run. This average value has so far been applied in the data evaluation of all GNO runs. A more detailed evaluation of the isotopic data is going on and samples from the other GNO runs will be measured in order to establish a comprehensive picture. Final corrections for each individual run will be possible when all MS data become available and the consistency of all GNO-measurements in this respect is ensured (within about one year from now). This is for scientific completeness. The maximum possible effective changes due to these refinements will under all circumstances be minor ( < 1 % ) with respect to the present overall result (the SNU-rate).

## 2.2 $^{69}$Ge counter calibration

The major contribution to the systematic errors of the GALLEX and GNO results so far (≈ 4 %) came from the insufficient knowledge of counter efficiencies (3.5 %). This is due to the fact that efficiencies for counters used in solar runs have not been measured directly because of contamination risk. Instead, they have indirectly been evaluated from measurements on other counters combined with a scaling procedure based on Monte Carlo simulations. However, some of the inputs needed in these MC simulations (counter volume, gas amplification curve) are not known with the accuracy that one can ambitiously desire.

In order to decrease this systematic error substantially, the GNO collaboration has developed a method that allows direct counter efficiency calibrations without introducing a major contamination risk [13]. $^{69}$Ge- (and some $^{71}$Ge-) activity is produced in the CN 7 MeV proton accelerator at the Legnaro INFN Laboratory (Italy) by the $^{69,71}$Ga (p,n)$^{69,71}$Ge reaction. $^{69}$Ge ($\tau$ = 56.3 h) decays by $\beta^+$ and EC. In the latter case, the signal produced in the counter is indistinguishable from the one of $^{71}$Ge decay, however coincident $\gamma$-rays are also emitted.

The irradiated Ga$_2$O$_3$ is transported to Heidelberg. Then, the radioactive Ge is converted to GeH$_4$ and filled into the counters to be calibrated (usually 2 or 3 counters in one calibration run). The absolute efficiency is determined by measuring the count rate in the proportional counter in coincidence with the 1106 keV $\gamma$ ray emitted in the EC decay of $^{69}$Ge, detected by a 9"×9" well-type NaI detector. Data have been recorded with a conventional MCA system. From April 2002 onwards a VME system operating in list mode was used in addition. The advantage of the latter is the fact that the full information about each event is recorded. This allows optimizing the $\gamma$-ray coincidence conditions off-line, instead of having them fixed at the time of measurement.

Six (p,n) irradiations and subsequent $^{69}$Ge calibration runs have been performed and a total of 12 GNO counters have been calibrated in these runs. While the statistical errors of the deduced efficiencies are always small (< 1 %), the limiting factor of the total error is due to systematic effects (corrections for $\beta^+$ contributions, the NaI background and instabilities of the electronics). The resulting total errors came out between 0.8 and 1.4 % (average 1.1 %). This constitutes a substantial reduction, as anticipated (compare columns 9 and 10 in Table 1). For 51 out of 58 GNO runs we have used counters that have been calibrated absolutely as described. Since the new efficiency determinations include also counters used in GNO I, slight changes of the data published in [1] result. This is one of the reasons why we have re-listed the GNO I results in Table 1. The absolute counter calibration program is going on, we plan to calibrate all counters (still alive) that have been used in relevant runs.

The efficiencies of the GNO counters depend only weakly on the gas filling (Xe/GeH$_4$ composition and pressure). Since the absolute efficiency measurements in Heidelberg have been performed with counter fillings close to the standard fillings in GNO runs, the efficiency values derived in Heidelberg are directly applicable to GNO solar runs.

## 2.3 Radon recognition and reduction

An elaborate radon test was performed with a modified counter containing a Ra source. The aim was to improve the characterization of radon events in the GNO proportional counters. The recordings lasted from May 1999 through March 2001 (1.8 years of counting time). After this long-lasting measurement, the emanation valve was closed and the intrinsic background of the counter was measured for 2.0 years. Using these data, we re-evaluated the inefficiency of the radon cut. The result is consistent with zero and the (2$\sigma$) upper value is 7.3 %. This replaces (for GNO)[1] the formerly determined value used in GALLEX, (9 $\pm$ 5) %.

Systematic tests on the synthesis line concerning GeCl$_4$ handling and the filling of GeH$_4$ into the counters have shown that the main sources of Rn contamination have been the mercury diffusion pump with its liquid nitrogen cold trap and the Xe storage vessel (Xe is admixed to GeH$_4$ to manufacture the counting gas).

The pump system was replaced by a turbo molecular pump in combination with a cold trap at -50 C°. The compression ratio for Rn is now increased by a factor of $\approx 10^6$ compared to the former case in which the diffusion pump was used. Furthermore, the storage vessel for Xe made of ©Jenaer Glass was replaced by a Kovar glass vessel which has a five times smaller concentration of $^{226}$Ra by weight.

---

[1] For GALLEX, the application of the more precise value would not be straightforward due to changes in electronics and some specific algorithms that have occurred between GALLEX and GNO data taking.

The mean level of identified $^{222}$Rn atoms per run during the GALLEX experiment was 4.5. Eleven runs were performed after the changes described above. Two runs with unusual technical problems yielded 3 atoms and 4 atoms of $^{222}$Rn, respectively. In the other runs we found 3 times one and 6 times zero $^{222}$Rn atoms. If the runs with technical problems are omitted, the mean value is now reduced to 0.33 $^{222}$Rn atoms per run.

**2.4  Electronics and noise reduction**

Analog and digital electronics, power supplies and data acquisition system have been completely renewed and reorganized after the accomplishment of GALLEX data taking. The analog bandwidth of the system has been increased to 300 MHz, the typical RMS noise is 2.8 mV. Due to these improvements and to a thorough screening of counters used in solar runs, the background in GNO is 0.06 counts per day in the relevant windows, corresponding to a 40 % background reduction compared to GALLEX.

In addition, a novel signal acquisition technique recently developed by us is the simultaneous recording of both, anode and cathode current from proportional counters [14]. The signal/noise ratio is increased by about 40 %. This is beneficial in particular for small signals. Furthermore, with this technique signals from inside the counters can be separated from signals picked up from the environment. For practical application, the counter boxes require mechanical modifications. So far, a prototype has been successfully applied in a blank run and in a solar run.

**2.5  GNO blank runs**

In addition to solar runs, one-day-exposure blank runs were also performed regularly in order to verify the absence of any artifact or systematics related to the target. During the period of operation of GNO, 12 blank runs ("BL") were successfully performed. The results are summarized in Table 3. The absence of spurious effects or unknown background is confirmed by the fact that the small excess of $^{71}$Ge counts in the blanks is consistent with the neutrino-induced production rate during the short exposure and the carry-over from the previous solar run.

**3. Data evaluation and counting results**

A random sampling of counter types was applied for the measurements (see Table 1, 8$^{th}$ column). For the selection of the $^{71}$Ge events we have subjected the counting data to our new neural network pulse shape analysis (NNPSA) [15] and to a subsequent maximum likelihood analysis [16].

**Table 3:** GNO blank runs

| Type[a] | Run# | Exposure time Start | Exposure time End | Duration [d] | Label[b] | Counting time [d] | Excess counts |
|---|---|---|---|---|---|---|---|
| BL1 | EX23 | 12/01/00 | 13/01/00 | 1 | SC138 | 164.5 | 0.0 |
| BL2 | EX27 | 05/04/00 | 06/04/00 | 1 | FC102 | 165.5 | 0.0 |
| BL3 | EX31 | 28/06/00 | 29/06/00 | 1 | FC126 | 164.1 | 0.0 |
| BL4 | EX35 | 20/09/00 | 21/09/00 | 1 | FC093 | 165.6 | 0.0 |
| BL5 | EX39 | 12/12/00 | 13/12/00 | 1 | Fe039 | 165.3 | 0.0 |
| BL6 | EX43 | 07/03/01 | 08/03/01 | 1 | Si108 | 164.8 | 0.0 |
| BL7 | EX47 | 30/05/01 | 31/05/01 | 1 | SC156 | 160.9 | 0.0 |
| BL8 | EX52 | 19/09/01 | 20/09/01 | 1 | Si119 | 166.9 | 4.0 |
| BL9 | EX56 | 12/12/01 | 13/12/01 | 1 | Fe039 | 172.6 | 1.8 |
| BL10 | EX60 | 06/03/02 | 07/03/02 | 1 | FC174 | 140.9 | 1.6 |
| BL11 | EX70 | 20/11/02 | 21/11/02 | 1 | Fe112 | 270.0 | 1.4 |
| BL12 | EX74 | 12/02/03 | 13/02/03 | 1 | FC102 | 189.1 | 0.0 |
| Average per run (combined analysis) | | | | | | | 0.43 ± 0.42 |

[a] BL = blank run.
[b] Counters have either iron (Fe) or silicon (Si) cathode. SC = silicon counter with shaped cathode, FC = iron counter with shaped cathode.

Pulses are first selected according to their amplitude (L or K peak), following the procedure described in [1]. Those passing the energy cut are then subject to a pulse shape selection. This procedure is logically divided into two steps: (a) fit of the pulse; (b) event selection through neural network. The pulses are fitted with a semi-empirical function obtained by the numerical convolution of the ideal pulse produced by a cylindrical proportional counter with the experimental response function of the electronic chain and a Gaussian charge collection function. The latter models a distributed deposition of energy. The information obtained by the $\chi^2$-fit of the pulses is then fed to a feed-forward 3-level neural network (NN) in order to distinguish between genuine $^{71}$Ge decays and background events. Since L and K captures create two distinct classes of events, two NNs are used, allowing to take the occurrence of double-ionization pulses after K-capture decays of $^{71}$Ge into account. Details about the training of the NN and about the validation of the procedure can be found in [15].

We have verified that the NNPSA provides a better noise rejection than the previously used rise time selection method, especially in the L energy window, at similar $^{71}$Ge acceptance efficiency ($\geq$ 93 %). In particular, NNPSA is able to recognize some of the fast-rising background pulses that otherwise would have been accepted on the basis of their rise time alone, and to treat the double-ionization events in the K window in the proper way. All the results obtained with both methods are consistent, but with lower background levels for NNPSA. Moreover, since other discriminating parameters are taken into account in addition to rise time, the NN-based approach is more robust and effective than the previous one.

The total GNO neutrino exposure time is 1687 days. During this time, the maximum likelihood analysis identifies a total of 258 decaying $^{71}$Ge atoms (131 L, 127 K), 239 of them (or 4.1 per run) due to solar neutrinos. The mean $^{71}$Ge count rate per run and counter at the start of counting is $\approx$ 0.27 counts per day. This may be compared with the *time independent* counter background in the acceptance windows.

**Table 4:** Background rates for counters used in GNO (0.5 keV ≤ E ≤ 15 keV)

| Counter | SR | BL | L-window fast, > 0.5 keV | K-window fast | L-window NNPSA | K-window NNPSA | Integral all |
|---|---|---|---|---|---|---|---|
| SC135 | 1 | - | 0.018 | 0.005 | 0.020 | 0.014 | 0.38 |
| SC136 | 7 | - | 0.031 | 0.012 | 0.033 | 0.018 | 0.32 |
| SC138 | 3 | 1 | 0.029 | 0.015 | 0.029 | 0.010 | 0.32 |
| SC139 | 4 | - | 0.015 | 0.010 | 0.018 | 0.010 | 0.25 |
| SC150 | 2 | - | 0.046 | 0.021 | 0.038 | 0.032 | 0.34 |
| SC151 | 2 | - | 0.036 | 0.023 | 0.052 | 0.023 | 0.37 |
| SC156 | 1 | 1 | 0.043 | 0.009 | 0.043 | 0.003 | 0.35 |
| Si106 | 5 | - | 0.034 | 0.017 | 0.029 | 0.012 | 0.41 |
| Si108 | 5 | 1 | 0.029 | 0.018 | 0.024 | 0.015 | 0.30 |
| Si113 | 2 | - | 0.047 | 0.037 | 0.032 | 0.029 | 0.38 |
| Si114 | 1 | - | ≤ 0.12 | ≤ 0.06 | ≤ 0.12 | ≤ 0.06 | 0.55 |
| Si119 | - | 1 | 0.017 | 0.036 | 0.010 | 0.023 | 0.64 |
| FC093 | 6 | 1 | 0.045 | 0.034 | 0.043 | 0.028 | 0.60 |
| FC102 | 4 | 2 | 0.049 | 0.025 | 0.059 | 0.030 | 0.47 |
| FC126 | 3 | 1 | 0.038 | 0.024 | 0.031 | 0.021 | 0.38 |
| FC174 | 4 | 1 | 0.038 | 0.024 | 0.034 | 0.022 | 0.52 |
| Fe039 | 4 | 2 | 0.029 | 0.013 | 0.020 | 0.013 | 0.49 |
| Fe043 | 1 | - | 0.131 | 0.087 | 0.084 | 0.083 | 0.88 |
| Fe112 | 2 | 1 | 0.047 | 0.027 | 0.050 | 0.018 | 0.48 |
| Fe118 | 1 | - | 0.017 | 0.027 | 0.017 | 0.028 | 0.46 |
| mean | | | 0.040 | 0.025 | 0.036 | 0.023 | 0.45 |

Sum of fast K + L : 0.065 cpd ;   Sum of NNPSA-selected K + L : 0.059 cpd

For the K- and L energy windows, the quoted rates apply to the fast-rising or NNPSA-selected background pulses that can mimic $^{71}$Ge pulses. The integral rate includes all the pulses in the whole energy range 0.5 keV ≤ E ≤ 15 keV.

**Table 5:** Results for individual solar neutrino runs in GNO III, GNO II, and GNO I

| Solar run | Result [SNU] | Solar run | Result [SNU] | Solar run | Result [SNU] |
|---|---|---|---|---|---|
| **GNO III** | | **GNO II** | | **GNO I** | |
| SR44 | $93 \pm ^{48}_{39}$ | SR20 | $52 \pm ^{44}_{34}$ | SR01 | $76 \pm ^{46}_{37}$ |
| SR45 | $95 \pm ^{52}_{42}$ | SR21 | $49 \pm ^{57}_{42}$ | SR02 | $50 \pm ^{47}_{35}$ |
| SR46 | $34 \pm ^{36}_{28}$ | SR22 | $-1 \pm ^{34}_{20}$ | SR03 | $97 \pm ^{64}_{51}$ |
| SR47 | $101 \pm ^{57}_{43}$ | SR23 | $125 \pm ^{53}_{43}$ | SR04 | $67 \pm ^{50}_{39}$ |
| SR48 | $64 \pm ^{51}_{37}$ | SR24 | $80 \pm ^{48}_{36}$ | SR05 | $-55 \pm ^{40}_{35}$ |
| SR49 | $-67 \pm ^{49}_{10}$ | SR25 | $64 \pm ^{57}_{44}$ | SR06 | $9 \pm ^{44}_{34}$ |
| SR50 | $7 \pm ^{47}_{37}$ | SR26 | $101 \pm ^{50}_{40}$ | SR07 | $120 \pm ^{55}_{46}$ |
| SR51 | $70 \pm ^{52}_{43}$ | SR27 | $40 \pm ^{50}_{36}$ | SR08 | $-30 \pm ^{34}_{25}$ |
| SR52 | $28 \pm ^{31}_{22}$ | SR28 | $62 \pm ^{40}_{29}$ | SR09 | $86 \pm ^{49}_{41}$ |
| SR53 | $50 \pm ^{45}_{32}$ | SR29 | $2 \pm ^{45}_{30}$ | SR10 | $141 \pm ^{58}_{46}$ |
| SR54 | $3 \pm ^{26}_{23}$ | SR30 | $109 \pm ^{51}_{42}$ | SR11 | $79 \pm ^{48}_{36}$ |
| SR55 | $98 \pm ^{48}_{38}$ | SR31 | $106 \pm ^{57}_{46}$ | SR12 | $36 \pm ^{69}_{40}$ |
| SR56 | $70 \pm ^{48}_{36}$ | SR32 | $69 \pm ^{40}_{30}$ | SR13 | $117 \pm ^{56}_{47}$ |
| SR57 | $91 \pm ^{44}_{35}$ | SR33 | $30 \pm ^{37}_{26}$ | SR14 | $129 \pm ^{72}_{58}$ |
| SR58 | $0 \pm ^{24}_{23}$ | SR34 | $160 \pm ^{58}_{50}$ | SR15 | $46 \pm ^{38}_{28}$ |
| | | SR35 | $63 \pm ^{39}_{29}$ | SR16 | $52 \pm ^{42}_{33}$ |
| | | SR36 | $19 \pm ^{46}_{21}$ | SR17 | $38 \pm ^{32}_{24}$ |
| | | SR37 | $63 \pm ^{53}_{41}$ | SR18 | $54 \pm ^{45}_{35}$ |
| | | SR38 | $61 \pm ^{53}_{41}$ | SR19 | $68 \pm ^{47}_{36}$ |
| | | SR39 | $61 \pm ^{49}_{60}$ | | |
| | | SR40 | $0 \pm ^{31}_{21}$ | | |
| | | SR41 | $93 \pm ^{47}_{38}$ | | |
| | | SR42 | $33 \pm ^{32}_{21}$ | | |
| | | SR43 | $107 \pm ^{52}_{41}$ | | |
| all: L only | $61.7 \pm ^{16.8}_{15.3}$ | all: L only | $68.5 \pm ^{14.3}_{13.3}$ | all: L only | $73.4 \pm ^{16.5}_{15.2}$ |
| all: K only | $49.7 \pm ^{12.4}_{11.2}$ | all: K only | $65.7 \pm ^{11.3}_{10.4}$ | all: K only | $60.1 \pm ^{13.0}_{12.0}$ |
| **all: GNO III** | $\mathbf{54.3 \pm ^{9.9}_{9.3}}$ | **all: GNO II** | $\mathbf{66.8 \pm ^{8.8}_{8.3}}$ | **all: GNO I** | $\mathbf{65.6 \pm ^{10.2}_{9.6}}$ |

Table 4 gives a summary of the measured backgrounds for the counters used in GNO, both with the rise time and the NNPSA selection. It is on average as low as $\approx 0.06$ counts per day (Table 4). For a comprehensive documentation of our counter backgrounds, which is also of general interest concerning the frontiers of 'Low-Level' counting, see [17].

The individual run results for the net solar production rates of $^{71}$Ge (based on the counts in the K and L energy- and neural network acceptance region) are plotted in Figure 1 and listed in Table 5, *after* subtraction of 4.55 SNU for side reactions as quoted in [1] and the correction for the annual modulation. Also listed in Table 5 are the combined results for the operating periods GNO III, GNO II and GNO I, respectively.

**Figure 1**: GNO single run results (SR1-SR58, see Table 1). Plotted is the net solar neutrino production rate in SNU after subtraction of side reaction contributions (see text). Error bars are ±1σ, statistical only.

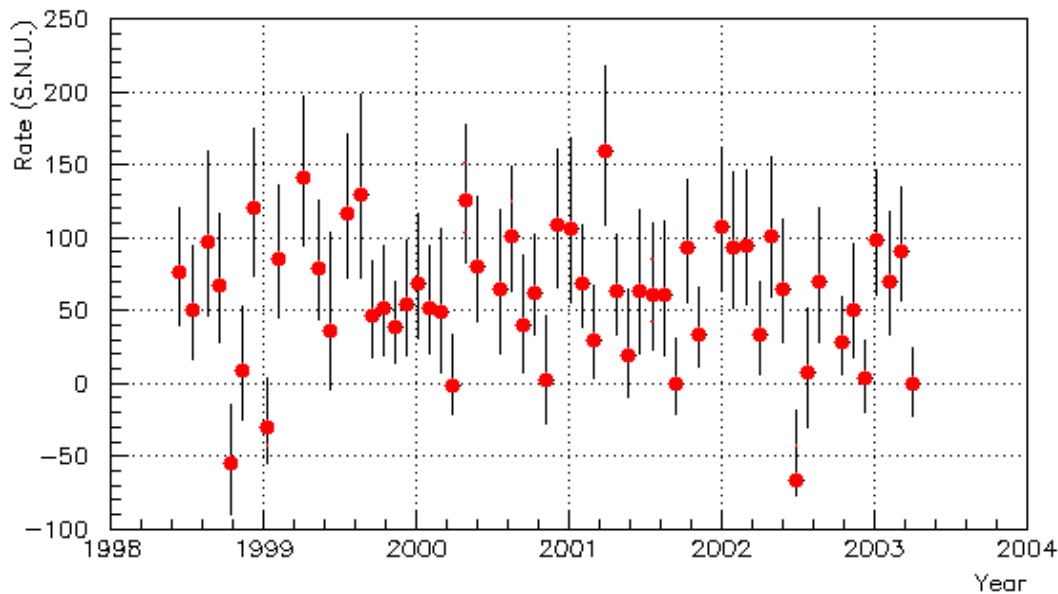

The systematic errors are specified in Table 6. The total systematic 1σ error is 2.5 SNU. It is still dominated by the error of the counting efficiencies, however it is substantially lower than in GALLEX (4.5 SNU).

**Table 6:** Systematic errors in GNO (all 58 runs)

|  | [%] | [SNU] |
|---|---|---|
| Target size | 0.8 | 0.5 |
| Chemical yield | 2.1 | 1.3 |
| Counting efficiency (energy cuts) | 2.2 | 1.4 |
| Counting efficiency (pulse shape cuts) | 1.3 | 0.8 |
| sub-total | 3.4 | 2.1 |
| side reactions subtraction error | 2.1 | 1.3 |
| **Total systematic errors** | **4.0** | **2.5** |

With 58 GNO runs at hand, the decay characteristics of $^{71}$Ge are very well reflected in the data. The energy spectrum of all fast counts that occurred in GNO runs during the first 50 days (3 mean lives of $^{71}$Ge) is shown in Figure 2. It clearly identifies the $^{71}$Ge electron capture spectrum with its peaks at 1 keV (relaxation of L-shell vacancy) and 10 keV (K-shell).

**Figure 2**: GNO energy spectrum of the events in the solar runs SR1-SR58. The outer contour of the histogram encloses the fast counts that occurred during the first 50 days, graphically superimposed on the background spectrum (shaded) recorded after the first 50 days (normalized).

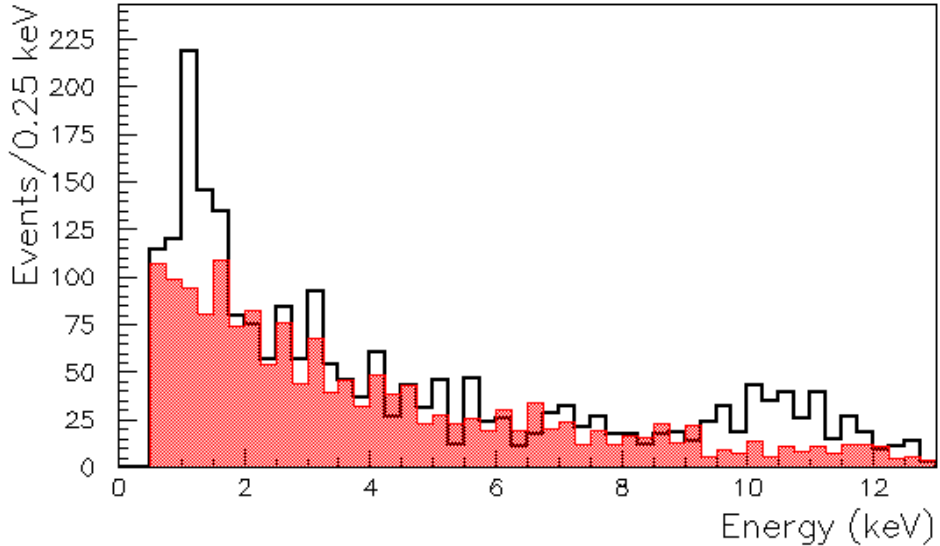

Similarly, if the $^{71}$Ge decay time $\tau_{71} = \tau(^{71}\text{Ge})$ is not fixed at its known value but treated as an additional free parameter in a combined Maximum Likelihood analysis of all GNO runs, a mean-life of $\tau_{71} = 16.6 \pm 2.1$ d (1$\sigma$) is obtained (Figure 3), in agreement with the known value of 16.49 d [18].

The combined net result for all GNO runs (after subtraction of 4.55 SNU for side reactions) is $62.9 \pm^{6.0}_{5.9}$ SNU (1$\sigma$, incl. syst.) (see sect.4, especially Table 7).

**Figure 3**: Counting rate of $^{71}$Ge candidates vs. time for the 58 solar runs of GNO. Dotted lines indicate $\pm 1\sigma$ envelopes.

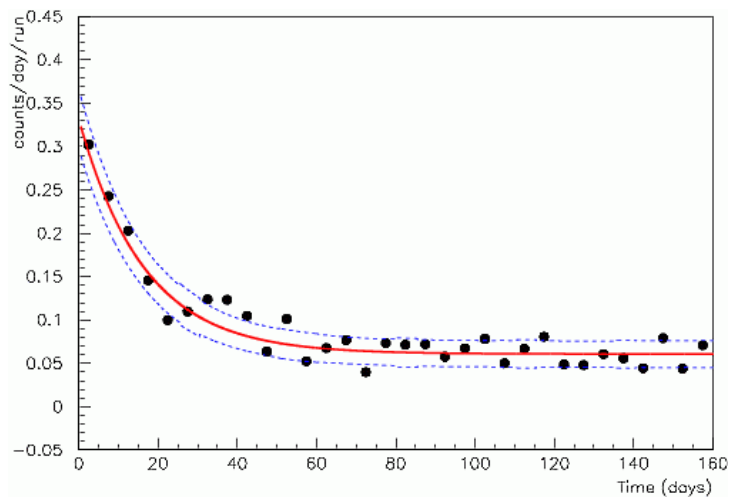

## 4. Joint analysis of GNO and GALLEX and an examination of a possible time variation of the neutrino signal

In Table 7 we compile the basic results for GNO, GALLEX, and for a combined analysis of GNO and GALLEX together. The joined GNO + GALLEX result after 123 solar runs is $69.3 \pm 4.1$ (stat.) $\pm 3.6$ (syst.) SNU (1$\sigma$) or $69.3 \pm 5.5$ SNU (1$\sigma$) with errors added in quadrature [19]. Our present joint result is now much more accurate, yet it retrospectively confirms the earlier GALLEX results which made the case to claim strong evidence for non-standard neutrino properties because the production rate predictions from the various standard solar models have always been much higher (120-140 SNU) than the measured rates [9]. The updated result of the SAGE experiment, $66.9 \pm^{5.3}_{5.0}$ SNU (1$\sigma$) [20], agrees well with our result.

**Table 7:** Results from GALLEX and GNO

|  | GNO | GALLEX | GNO + GALLEX |
|---|---|---|---|
| Time period | 05/20/98–04/09/03 | 05/14/91 – 01/23/97 [a] | 05/14/91– 04/09/2003 [b] |
| Net exposure time [d] | 1687 | 1594 | 3281 (8.98 yrs) |
| Number of runs | 58 | 65 | 123 |
| L only [SNU] | $68.2 \pm^{8.9}_{8.5}$ | $74.4 \pm 10$ | $70.9 \pm 6.6$ |
| K only [SNU] | $59.5 \pm^{6.9}_{6.6}$ | $79.5 \pm 8.2$ | $67.8 \pm 5.3$ |
| Result (all) [SNU] | $62.9 \pm^{5.5}_{5.3}$ stat. $\pm 2.5$ | $77.5 \pm 6.2$ stat. $\pm^{4.3}_{4.7}$ | $69.3 \pm 4.1$ stat. $\pm 3.6$ |
| Result (all) [SNU] [c] | $62.9 \pm^{6.0}_{5.9}$ incl. syst. | $77.5 \pm^{7.6}_{7.8}$ incl. syst. | $69.3 \pm 5.5$ incl. syst. |

[a] except periods of no recording: 5-8/92; 6-10/94, 11/95-2/96
[b] except periods of no recording: as before, + 2/97-5/98
[c] statistical and systematic errors combined in quadrature. Errors quoted are 1$\sigma$.

The results of all 123 individual GNO and GALLEX measurements of the neutrino capture rate are shown in Figure 4. Like in all of our previous papers, we estimate the gallium-solar neutrino interaction rate under the assumption of a neutrino flux constant in time. This assumption must be justified even though there are presently no attractive models that predict a time variable neutrino emission from the Sun. Observation of such a variation on a short (that is, non-secular) time scale would signal new and unexpected solar or neutrino physics. This cannot be excluded 'a priori'. Consequently, we analyze our (GNO+GALLEX) data for possible time variations during the ≈12-year period of data taking.

**Figure 4**: Single run results for GNO and GALLEX [7] during a full solar cycle. Plotted is the net solar neutrino production rate in SNU after subtraction of side reaction contributions (see text). Error bars are ±1σ, statistical only.

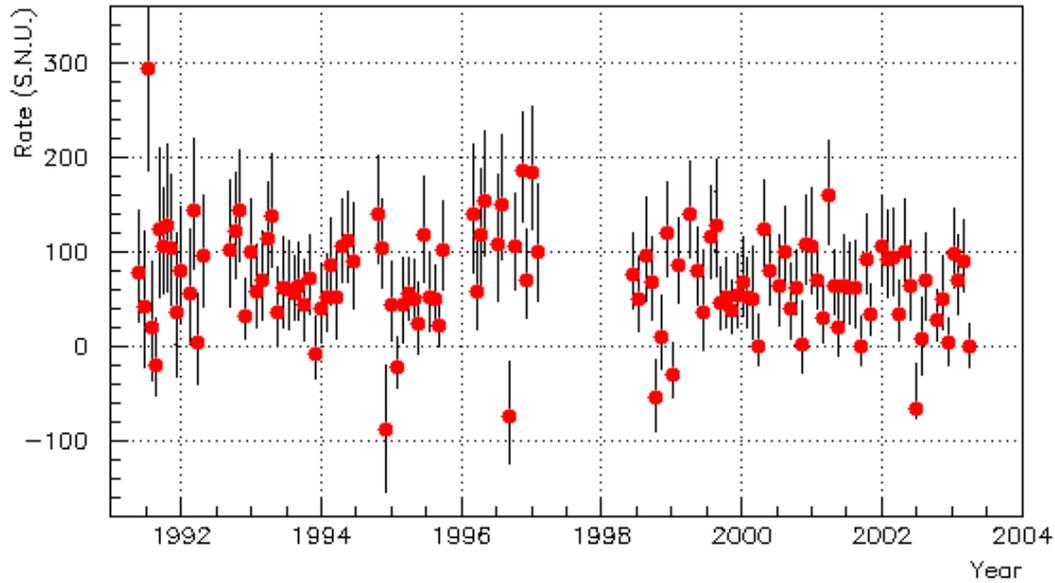

The scatter plots of the single run results for GNO, GALLEX and GALLEX+GNO are shown in Fig 5 (thick histograms). They are compatible with the Monte Carlo generated distributions of single run results for a constant production rate (62.9, 77.5 and 69.3 SNU respectively) under the typical solar run conditions (efficiencies, exposure time, etc.).

**Figure 5**: Distribution of GNO, GALLEX and GALLEX+GNO single run results (thick lines). They are superimposed with the corresponding Monte Carlo distributions (shaded histograms) obtained assuming a constant neutrino rate and taking into account the actual run conditions.
Confidence levels resulting from Neyman-Pearson and Kolmogorov-Smirnov tests between the Monte-Carlo and the experimental single run distributions are, respectively, 88 % and 82 % for GNO ($\chi^2$/ndf = 8.12/14); 6 % and 38 % for GALLEX ($\chi^2$/ndf = 25.6/16); 34 % and 51 % for GNO + GALLEX ($\chi^2$/ndf = 18.8/17).

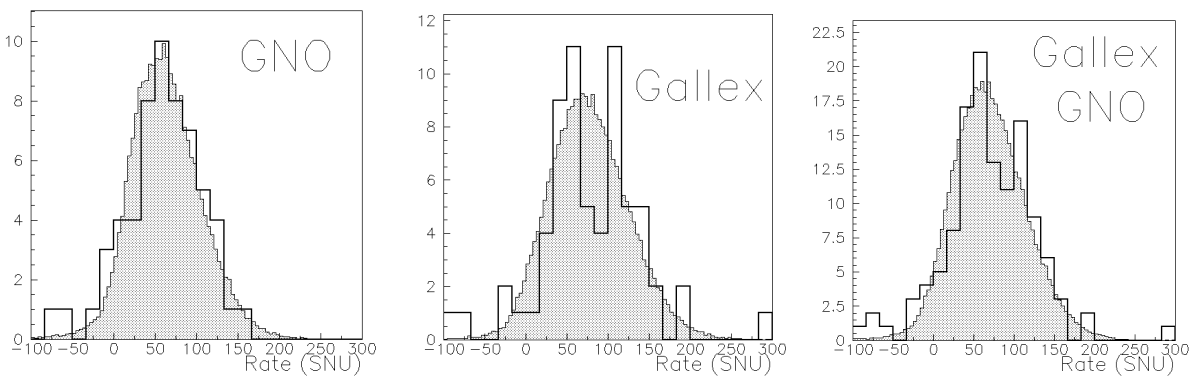

As can be seen from Fig. 5, the single run distributions can be approximated by quasi-Gaussian distributions having similar relative spread; this is due to the fact that the experimental conditions (exposure time, chemical yield, counting efficiencies..) are very similar for all GALLEX and GNO runs.

To test the null hypothesis of a capture rate constant in time we have used two different approaches:

- Application of the maximum likelihood ratio test (see [5] for details). The resulting goodness of fit confidence levels are 25.1 % for GNO, 24.2 % for GALLEX, and 5.6 % for GALLEX+GNO.
- Fits of the results of the seven periods GNO III, II, I and GALLEX IV, III, II, I with (arbitrary) time-varying functions, e.g. $f = a + bt$, and compare the results with the null hypothesis. The resulting confidence levels for these options do not differ significantly (see Fig. 6).

We conclude that the results are consistent with a flat behaviour; however a weak time dependency (of unknown origin) is not excluded.

**Figure 6**: GNO and GALLEX group results vs. time of measurement. The dotted lines indicate the fits for $f = a = $ const. (horizontal, $a = 69.3 \pm 4.1$ SNU, $\chi^2$/ndf = 13.2/6 , C.L. = 4 %) and for $f = a + bt$ (inclined, $f = [ (82 \pm 10) + ( -1.7 \pm 1.1) \, t \, yr^{-1}) ]$ SNU, $\chi^2$/ndf = 10.8/5 , C.L. = 5.5 %).

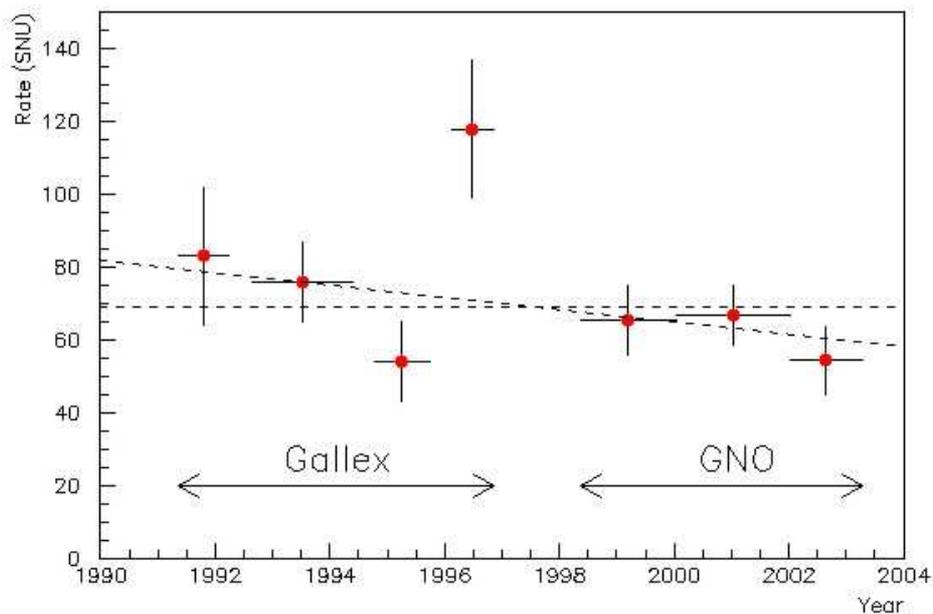

We have analyzed the GALLEX and GNO data relative to a correlation with the seasonal Earth-Sun distance variation. The 123 solar runs have been divided into 6 about equally populated bins of similar heliocentric distance d (Figure 7). The fit assuming a solar neutrino rate constant in time and affected only by the $1/d^2$ geometrical modulation yields a confidence level of 69 % ($\chi^2 = 3.0$ with 5 d.o.f.).

The difference between rates of the solar runs performed in winter time W (defined as perihelion $\pm 3$ months) and in summer time S is $\Delta(W-S) = -7.6 \pm 8.4$ SNU (the value expected from the $1/d^2$ modulation only is + 2.3 SNU).

**Figure 7**: GNO/GALLEX signal vs. heliocentric distance of the Earth. The straight line indicates the expected flux variation due to purely geometrical ($1/d^2$) effects.

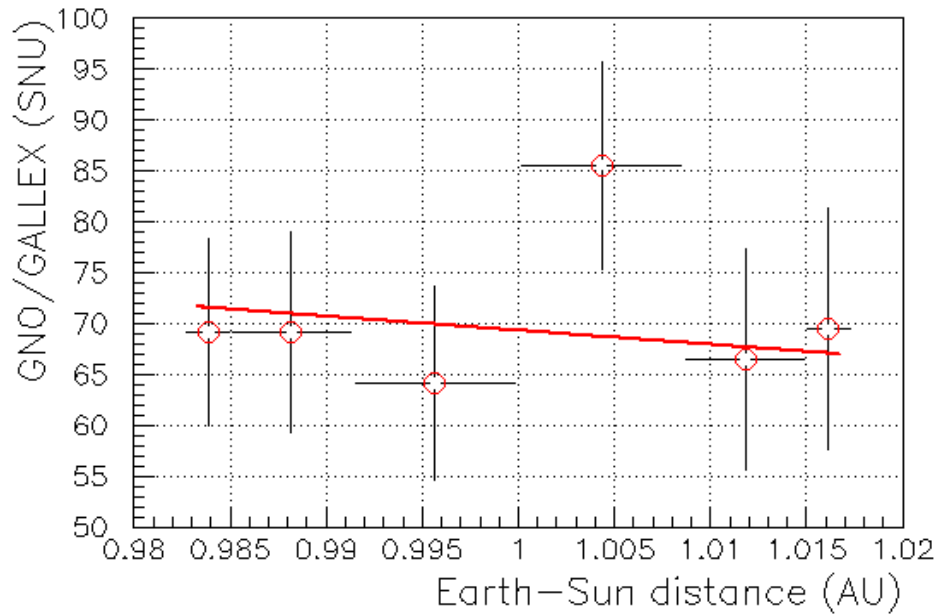

The finding that the solar data are consistent with a production rate constant in time does not invalidate other hypotheses that might give similar or even better (short) time dependent fits [21,22]. Respective searches for such time modulations in the GNO and GALLEX data [23] yield no statistically significant periodicity within the Lomb-Scargle [24,25] and maximum likelihood algorithms.

The sensitivity of any time series analysis of Ga radiochemical data is limited by the integration of the signal during the exposure time and by the large statistical error of the single runs. However, sinusoidal modulations with low frequency ($< 11$ yr$^{-1}$) and large amplitude ($> 60$ %) are already excluded at 90 % C.L. (see the exclusion plot, Fig. 8).

**Figure 8:** 90 % CL exclusion plot in the frequency/amplitude plane from the analysis of GALLEX and GNO data [23].

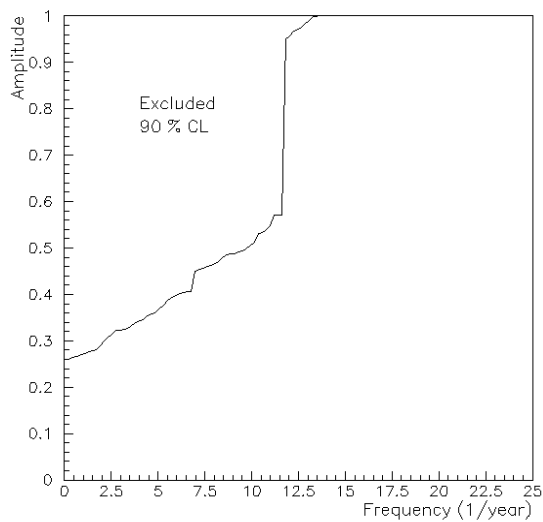

The results of 12 years of solar observations call for a continuous monitoring of the low energy solar neutrino flux with ever increasing sensitivity over very long time periods, in a spirit comparable to the motivations for a continuous observation of solar surface phenomena.

## 5. Implications
### 5.1 Particle physics context

Concerning both particle physics as well as astrophysics, low-energy solar neutrino experiments will be of high importance also in the years to come (e.g., [26]). Global fits of all solar and reactor neutrino data have established that the deficits observed in the signals of all experiments are caused by neutrino oscillations with parameters in the Large Mixing Angle (LMA) region (see e.g. [27] (SNO), [40] (Kamland), [26] (analysis)). If the LMA(MSW) oscillation solution is correct, the basic oscillation mechanism changes at a neutrino energy of about 2 MeV from the MSW matter mechanism (above 2 MeV) to the vacuum oscillation mechanism (below 2 MeV). This transition has not yet been checked experimentally in a model-independent way.

The gallium result fits well in the oscillation scenario. If one subtracts the $^8$B neutrino contribution as measured by SNO [27] from the gallium signal and calculates the suppression factor P with respect to the BP04 SSM [28] for sub-MeV neutrinos (pp and $^7$Be), the result is P = 0.556 ± 0.071. Assuming vacuum oscillations, P is given by: P = 1 - 0.5 $\sin^2(2\theta)$. This yields $\theta = 35.2^{+9.8}_{-5.4}$ degrees. Although such an estimate is quite approximate, it is in good agreement with the latest determination (32.0 ± 1.6 degrees [26]), which essentially comes from $^8$B neutrinos (i.e. matter effects dominate).

The GNO result is fully consistent both with the GALLEX result and with all other solar neutrino data if judged in the frame of the Standard Solar Model and the MSW scenario, where a gallium rate within a range of 70 ± 2 SNU is predicted.

In addition to the dominant LMA(MSW) conversion mechanism, the possible existence of sterile neutrinos and/or of flavor-changing neutrino matter interactions other than the MSW effect [29-31] can be investigated with low-energy solar neutrinos.

About 91 % of the total solar neutrino flux is expected to originate from low-energy pp-neutrinos. This fundamental prediction of the standard solar model has still to be tested experimentally by separate determinations of the solar neutrino fluxes of the pp, $^7$Be, and CNO reactions.

Considerable effort was devoted to the improvement of GNO sensitivity and to the reduction of the statistical as well as the systematic errors, e. g. the installation of new electronics, the introduction of neural network data analysis and developments towards cryogenic detectors for an improved $^{71}$Ge counting efficiency [3]. It will be for the next generation of large solar neutrino detectors to reach a sensitivity which not only will be sufficient for the determination of the individual solar neutrino fluxes but which will also enable investigations in other astrophysical fields via neutrino astronomy. Future low-energy neutrino detectors may very well allow studying physics beyond solar astronomy, for example supernovae and geophysics.

**5.2 Astrophysical context**

*5.2.1 Comparison of gallium experimental results with solar models*

Recent new experimental data on input parameters for solar model calculations led to improved predictions for neutrino fluxes and capture rates. In particular this refers to:

- new measurements of the solar surface composition [32]. The new determination of the C,N,O surface abundances changes the metal to hydrogen ratio Z/X from 0.0229 previously to 0.0176 now.
- new direct measurements of the $^7$Be $(p,\gamma)$ $^8$B cross section [33]. This led to a 15 % increase in the $^8$B neutrino flux (see Ref. [28], model BP04 vs. [34]).
- new measurement of the $^{14}$N$(p,\gamma)^{15}$O cross section [35] down to 70 keV.
- recalculation of $S_0$(pp) and $S_0$(hep).

The immediate consequences for the predicted neutrino fluxes (BP04$^+$ vs. BP04 in Ref. [28]) are reductions for $^8$B (9 %), $^{13}$N (30 %) and $^{15}$O (30 %) contributions (see Table 8).

**Table 8:** Predicted capture rate values for the radiochemical gallium and chlorine neutrino experiments according to various standard solar model calculations, (i) without and (ii) with neutrino oscillations (at global fit parameters).

| | **BP04** [28] | **BP04+** [28] | **Franec04+** [36] | **BP00** [34] | **Experim.** |
|---|---|---|---|---|---|
| Input: Z/X | 0.0229 | 0.0176 | 0.0176 | 0.0230 | |
| Outputs: Y | 0.243 | 0.2382 | 0.2390 | 0.2437 | |
| $R_{cz}$ | 0.715 | 0.726 | 0.730 | 0.715 | |
| $T_c$ (10$^7$K) | 1.572 | | 1.557 | 1.570 | |
| pp-flux (10$^9$/cm$^2$,s) | 59.4 ± 1 % | 59.9 ± 1 % | 60.19 ± 1 % | 59.5 ± 1 % | |
| pep-flux (10$^9$/cm$^2$,s) | 0.14 ± 2 % | 0.142 ± 1 % | 0.143 ± 2 % | 0.140 ± 2 % | |
| $^7$Be-flux (10$^9$/cm$^2$,s) | 4.86 ± 12 % | 4.65 ± 12 % | 4.62 ± 12 % | 4.77 ± 10 % | |
| $^8$B-flux (10$^7$/cm$^2$,s) | 0.579 ± 23 % | 0.526 ± 23 % | 0.487 ± 23 % | 0.505 ± $^{20}_{16}$ % | |
| $^{13}$N-flux (10$^9$/cm$^2$,s) | 0.571 ± $^{37}_{35}$ % | 0.406 ± $^{37}_{35}$ % | 0.230 ± $^{37}_{35}$ % | 0.548 ± $^{21}_{17}$ % | |
| $^{15}$O-flux (10$^9$/cm$^2$,s) | 0.503 ± $^{43}_{39}$ % | 0.354 ± $^{43}_{39}$ % | 0.173 ± $^{43}_{39}$ % | 0.480 ± $^{25}_{19}$ % | |
| $^{17}$F-flux (10$^7$/cm$^2$,s) | 0.591 ± 44 % | 0.397 ± 44 % | - | 0.563 ± 25 % | |
| (i) Cl$_{th}$ no osc. (SNU) | 8.5 ± 1.8 | 7.7 ± 1.6 | 7.0 ± 1.5 | 7.6 ± $^{1.3}_{1.1}$ | |
| (i) Ga$_{th}$ no osc. (SNU) | 131 ± $^{12}_{10}$ | 126 ± $^{12}_{10}$ | 122 ± $^{11}_{9}$ | 128 ± $^{9}_{7}$ | |
| (ii) Cl$_{th}$ incl. osc.(SNU)* | 3.2 ± 0.7 | 2.9 ± 0.6 | 2.7 ± 0.6 | 2.9 ± 0.4 | 2.56 ± 0.23 [37] |
| (ii) Ga$_{th}$ incl. osc. (SNU)* | 72 ± $^{6.6}_{5.5}$ | 69.6 ± $^{6.3}_{5.3}$ | 67.6 ± $^{6}_{5}$ | 70.9 ± $^{2.6}_{2.3}$ | 68.1 ± 3.8 |

* $\delta m^2 = 8.2 \cdot 10^{-5}$ (eV/c$^2$)$^2$ ; $\tan^2\theta = 0.39$

The updated model BP04$^+$ foresees a depth of the convective zone $R_{CZ}/R_0 = 0.726$. This is clearly in conflict with the very accurately measured helioseismological value of $0.713 \pm 0.001$. The change happns because once the surface heavy element composition is decreased, the radiative opacity and the central temperature will also decrease and the base of the convective zone is moving outward.

Due to this conflict, we prefer for the time being the *partially* updated model BP04 [28]. This model obeys neither the new value for Z/X nor the new value for $S_0[^{14}N(p,\gamma)^{15}O] = 1.77 \pm 0.2$ keVbarn which practically is one-half of the previous best estimate for this cross section. This would cut almost in half the $^{13}$N and $^{15}$O neutrino fluxes (see Table 8, column for Franec04$^+$).

In absolute terms, CNO nuclear reactions contribute 1.6 % to the solar luminosity in BP04 [28], however only 0.8 % in Franec04$^+$ [36]. We included Franec04$^+$ in Table 8 to illustrate the robustness of the neutrino capture rate predictions even under heavy modifications of the model calculations. Given the accuracy of both the gallium[2] and the chlorine experimental results it is impossible to distinguish between BP04 and Franec04$^+$, one can just notice that both the gallium and mainly the chlorine results are on the lower side of the predictions.

*5.2.2 Limits on the CNO cycle contributions to the solar luminosity*

Due to the very low energy threshold, gallium experiments are still the only source of experimental information about sub-MeV solar neutrinos (in particular about the fundamental pp-neutrinos). These experiments test the consistency of the (electromagnetic) solar luminosity with the observed neutrino fluxes. The luminosity constraint is defined by

$$L_{Sun} = \Sigma_i \Phi_i \cdot \alpha_i \qquad (1)$$

where $L_{Sun} = 8.53 \cdot 10^{11}$ MeV cm$^{-2}$s$^{-1}$ is the solar luminosity.

$\Phi_i$ and $\alpha_i$ are, respectively, the neutrino flux and the energy release in photons per emitted neutrino of type i ( i = pp, pep, $^7$Be, ...).

This assumes that nuclear fusion reactions are the only energy production mechanism inside the Sun. Apart from this basic assumption, the luminosity constraint does not depend on the solar model.

The fractional CNO luminosity is defined by:

$$\frac{L_{CNO}}{L_{Sun}} = \frac{\Phi_O \cdot \alpha_O + \Phi_N \cdot \alpha_N}{\Sigma_i \Phi_i \cdot \alpha_i} \qquad (2)$$

It can be calculated from the measured gallium rate

---

[2] The gallium result reported here is the combined value from GALLEX/GNO and SAGE, as commonly adopted in global fit analyses (e.g., [28]). Statistical and systematic errors are first added in quadrature for each single experiment, then the individual experimental results are combined with the usual statistical rules.

$$R^{Ga} = \sum_i \int \sigma_i(E)\varphi_i(E)P(E)dE \tag{3}$$

where $\sigma_i(E)$ is the neutrino capture cross section on Ga, $\varphi_i(E)$ is the differential flux of solar neutrinos of species $i$, and $P(E)$ is the electron neutrino survival probability. For this calculation we make the following assumptions:

- the $^8$B electron neutrino flux, and the electron neutrino survival probabilities are measured with a precision of the order of 12 % by SNO [27].
- the $^7$Be neutrino flux is as deduced in BP04 SSM [28], with an uncertainty of 12 %. It is not directly measured up to now.
- the neutrino flux ratios pep/pp and $^{13}$N/$^{15}$O are fixed from nuclear physics and kinematics with negligible uncertainties (see [26]).
- the neutrino capture cross section (and its uncertainty) on $^{71}$Ga is theoretically calculated as in [38].

With these assumptions we deduce from the gallium capture rate as measured by GALLEX/GNO (and applying the oscillation parameters $\delta m^2 = 8.2 \cdot 10^{-5}$ (eV/c$^2$)$^2$; $\tan^2\theta = 0.39$ ) the following upper limit:

$$L_{CNO}/L_{Sun} < 6.5 \% \ (3\sigma) \ (\text{best fit: } 0.8 \%). \tag{4}$$

**Figure 9**: Plot of the solar luminosity fraction due to CNO-cycle reactions versus the gallium neutrino capture rate. The underlying assumptions are discussed in the text. Contours are shown for the 1σ, 2σ, and 3σ limits that are allowed by the GALLEX/GNO experimental result on the gallium rate. The straight line is given by the luminosity constraint (see text).

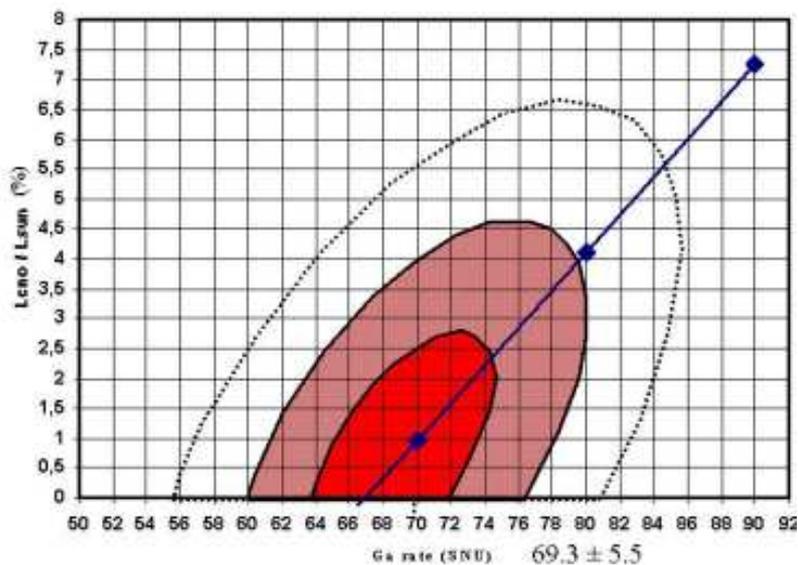

The result is represented graphically in Figure 9 in the plane $L_{CNO}/L_{Sun}$ vs. Ga rate. Plotted are the regions that are allowed at the 1σ, 2σ and 3σ sigma levels. The result is in good agreement with the predictions of the solar models, $L_{CNO} = 1.6 \pm 0.6$ %. This is a unique self-consistency test of the observed solar luminosity, the predicted neutrino fluxes, and the oscillation scenario.

We stress again that in order to obtain the above results we have assumed that the $^7$Be neutrino flux is as predicted from the SSM, with an uncertainty of 12 %. Hopefully, this flux will soon be directly measured by BOREXINO [39] and/or KAMLAND [40]. When this will be done, the gallium rate will become a completely model independent test of the solar neutrino luminosity. An upper limit on the CNO luminosity from all currently available solar neutrino and reactor anti-neutrino experimental data is discussed in [26].

____________________________________________________________

## Acknowledgements

We thank the GALLEX Collaboration for the permission to access the GALLEX raw data for a joint analysis in combination with GNO data. We are obliged to A. Bettini, former director of LNGS, for his competent and productive support and advice. We also want to acknowledge the help of L. Ioannucci and the staff of LNGS.